# An improved DNN-based spectral feature mapping that removes noise and reverberation for robust automatic speech recognition

*Juan Pablo Escudero[1], José Novoa[1], Rodrigo Mahu[1], Jorge Wuth[1], Fernando Huenupán[2], Richard Stern[3] and Néstor Becerra Yoma[1]*

[1] Speech Processing and Transmission Laboratory, Electrical Engineering Department, Universidad de Chile, Santiago, Chile.
[2] Department of Electrical Engineering, Universidad de la Frontera, Temuco, Chile.
[3] Department of Electrical and Computer Engineering and Language Technologies Institute, Carnegie Mellon University, Pittsburgh, USA.

nbecerra@ing.uchile.cl

## Abstract

Reverberation and additive noise have detrimental effects on the performance of automatic speech recognition systems. In this paper we explore the ability of a DNN-based spectral feature mapping to remove the effects of reverberation and additive noise. Experiments with the CHiME-2 database show that this DNN can achieve an average reduction in WER of 4.5%, when compared to the baseline system, at SNRs equal to -6 dB, -3 dB, 0 dB and 3 dB, and just 0.8% at greater SNRs of 6 dB and 9 dB. These results suggest that this DNN is more effective in removing additive noise than reverberation. To improve the DNN performance, we combine it with the weighted prediction error (WPE) method that shows a complementary behavior. While this combination provided a reduction in WER of approximately 11% when compared with the baseline, the observed improvement is not as great as that obtained using WPE alone. However, modifications to the DNN training process were applied and an average reduction in WER equal to 18.3% was achieved when compared with the baseline system. Furthermore, the improved DNN combined with WPE achieves a reduction in WER of 7.9% when compared with WPE alone.

## 1. Introduction

In indoor environment, the reverberation and additive noise have detrimental effects on the performance of ASR systems [1]. Therefore, the challenge is to implement methods capable of reducing or eliminating the reverberation and the additive noise in speech signals [2] [3]. The methods are divided in three classes depending on where they are implemented in an ASR system: front-end; back-end; and, speech preprocessing [4]. The preprocessing methods are also known as speech enhancement (SE).

The use of neural networks (NN) for SE or speech distortion removal is not new. In [5] a multilayer perceptron NN inspired in the Lateral Inhibition process was proposed to cancel additive noise. Some DNN methods are focused on learning clean version of MFCC features using either de-noising autoencoders [6] [7] or recurrent neural network (RNN) [8]. Furthermore, spectral subtraction and a RNN based de-noising autoencoder were implemented to learn Mel filterbank features from the noise [9].

Particularly, in [10] a DNN-based spectral feature mapping (named SFM-DNN here and after) was proposed. SFM-DNN has the capability to map spectral features on Mel filterbank features, while in the process the additive noise and reverberation are canceled or reduced. However, the resulting SFM-DNN removes additive noise more effectively than reverberation.

SFM-DNN was built with a multilayer perceptron composed of two hidden layers, each one with 2048 units, and one output layer with 40 units. The activation functions of both hidden and output layer correspond to the sigmoid function. The input is fed with spectrograms extracted from corrupted voice samples. The reference used to train de DNN were the log Mel filterbank features extracted from the corresponding clean speech samples. The input features were normalized to zero mean and unit variance over all feature vectors in the training set. Since temporal dynamics incorporates rich information for speech, 10 neighborhood frames were included to the current frame. The reference features were normalized in a range of [0,1]. The DNN was trained using the backpropagation with mini-batch stochastic gradient descendent algorithm. The adaptive gradient descendent was used as optimization technique and the cost function was based on the mean square error.

WPE is a widely employed enhancement method for de-reverberation corresponds [11]. WPE has led to significant improvements in ASR accuracy under different reverberant conditions [2]. This method is focused on the reverberation suppression by means of the iterative estimation of linear regression filter coefficients. The technique performs the coefficient estimation using the short time Fourier transform (STFT). At the end of the process, a reverberation free speech time waveform is obtained.

In this paper we improve the SFM-DNN performance by making use of WPE. Furthermore, we introduced modifications to the SFM-DNN training process to improve its effectiveness in distortion removal. As a result, a final WER equal to 13.2% and a relative reduction in WER as high as 18.3% when compared to the baseline system with noisy training were achieved. The results reported in this paper are very competitive with those published elsewhere using the same database that was presently employed [12] [10] [13] [14] [15]. Also, the reduction in WER when compared to the original SFM-DNN was 15.1% In section 2 we present and justify the combination of SFM-DNN with WPE. Section 3 describes the modifications and improvements incorporated to the SFM-DNN training process. The discussions and conclusions are presented in sections 4 and 5, respectively.

Table 1: *WER (%) corresponding to the validation results of the original SFM-DNN using CHiME-2 noisy training with clean alignment [10].*

|      | System Training |              |                 |                  |
|------|-----------------|--------------|-----------------|------------------|
| Test | Noisy [10]      | SFM-DNN [10] | Noisy (Local)   | SFM-DNN (Local)  |
| -6 dB | 28.2           | 28.0         | 26.8            | 24.8             |
| -3 dB | 20.7           | 19.9         | 20.6            | 19.5             |
| 0 dB  | 16.3           | 14.8         | 16.2            | 15.7             |
| 3 dB  | 13.1           | 11.9         | 13.2            | 13.0             |
| 6 dB  | 9.5            | 10.2         | 10.6            | 10.6             |
| 9 dB  | 9.1            | 8.9          | 9.7             | 9.8              |
| Avg.  | 16.1           | 15.6         | 16.2            | 15.6             |

## 2. Combining the DNN with WPE

First of all, we replicated the results reported in [10]. The experiment was performed using the CHiME-2 track 2 database [3]. DNN was implemented using the TensorFlow's Python API [16].

The ASR experiments were conducted using the Kaldi toolkit [17]. The ASR system is based on a DNN-HMM, which use as input 40-dimentional log Mel filterbank with deltas and delta-delta dynamic features from an 11-frame context window. The DNN-HMM was trained using the tied triphone state targets obtained from a clean GMM-HMM alignment. A realignment was performed using the trained DNN [18], then a retrain of the DNN was performed with the new alignment. The DNN-HMM was improved applying sMBR-based sequence training on the DNN [19].

Table 1 shows the WERs reported in [10] and the validated results with the local system. These results suggest that SFM-DNN has the ability to reduce the effect of additive noise rather than reverberation. For this reason, we could improve the SFM-DNN performance by combining it with a SE method focused on reverberation reduction, i.e. WPE.

The cascade configuration of enhancement methods deserves a brief discussion. For instance, in [20] it is discussed that the channel response can effectively be removed after the additive noise being cancelled by considering that the former is supposed time-invariant and the latter can be non-stationary. However, in the case considered here, the reverberation and additive noise can be both assumed non-stationary. Consequently, the order in the cascade removal would not seem relevant. However, in the case considered here, the cascade sequence order is constrained by the input and output of SFM-DNN. As a result, we proposed the scheme shown in Fig. 1.

Figure 2 shows the reductions in WER compared to the baseline system (noisy training, Table 1) provided by SFM-DNN and WPE methods vs. SNR. The SFM-DNN scheme provides a reduction in WER that decreases when SNR increases. Actually, at SNR equal to 9 dB SFM-DNN introduces a distortion. These results suggest that SFM-DNN cancels more effectively the additive noise rather than reverberation. In contrast, WPE provides a reduction in WER that is relatively constant up to SNR equal to 6 dB and then increases significantly when SNR is equal to 9 dB.

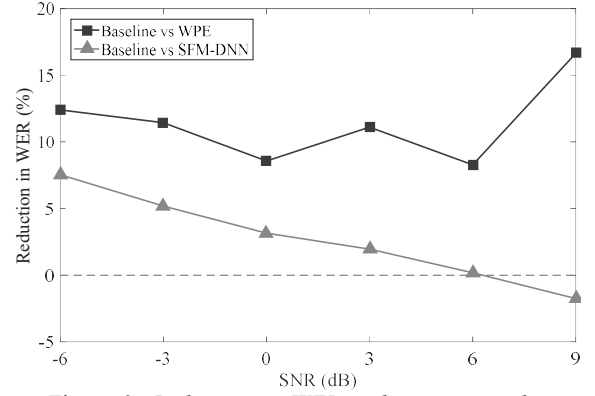

Figure 2: *Reduction in WER with respect to the Baseline, for WPE and SFM-DNN systems.*

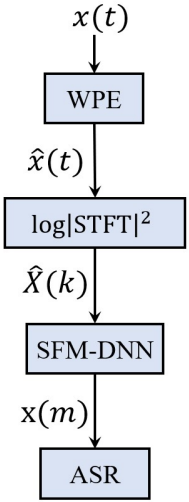

Figure 1: *Block diagram of the proposed cascade sequence of SE methods. $x(t)$, $\hat{x}(t)$, $\hat{X}(k)$ and $x(m)$ represent, respectively, the speech signal in the temporal domain, the free reverberation speech time waveform, log spectral magnitude and the resulting Mel filterbank enhanced speech signal.*

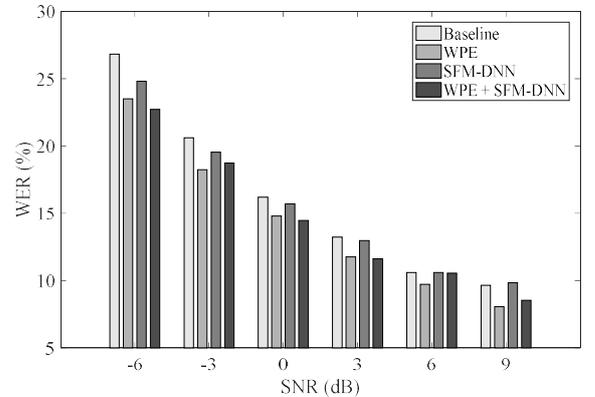

Figure 3: *WER (%) with the baseline, WPE, SFM-DNN and WPE+SFM-DNN systems.*

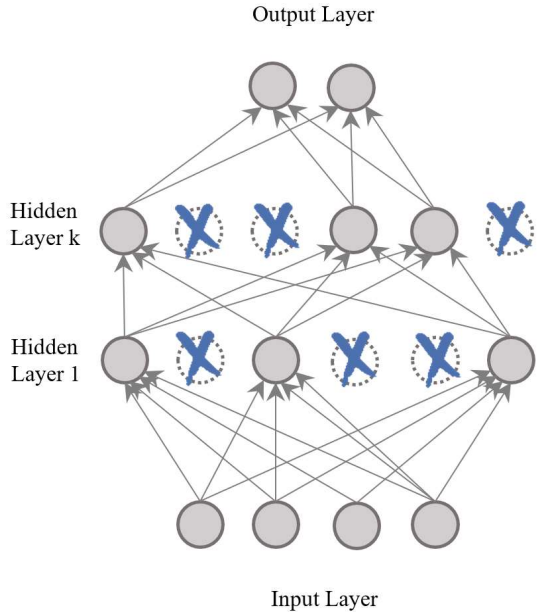

Figure 4: *Graphic description of the dropout method for a typical neural network architecture.*

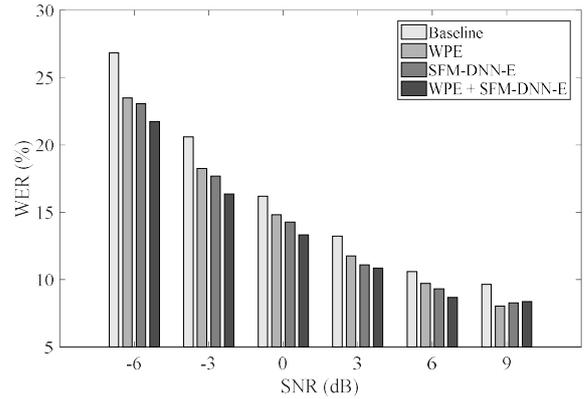

Figure 5: *WER (%) with the baseline, WPE, SFM-DNN-E and WPE+SFM-DNN-E systems.*

Figure 3 shows the results of ASR experiments obtained with the Baseline, WPE, SFM-DNN and WPE+SFM-DNN (according to Fig.1) systems. When compared to the baseline system, SFM-DNN and WPE+SFM-DNN provided reductions in WER equal to 7.3% and 10.8%, respectively, which in turn validates the proposed approach illustrated in Fig. 1. However, WPE led to a higher WER reduction (11.4%) than WPE+SFM-DNN. Consequently, while SFM-DNN was improved with the addition of WPE, the resulting WPE+SFM-DNN system is somehow limited by the performance of SFM-DNN. This result motivated us to improve the SFM-DNN effectiveness by incorporating modifications to the training process.

## 3. DNN training modifications and improvements

In order to improve the SFM-DNN effectiveness we explored the use of dropout, cross-validation and alternative input-reference normalization. Our reference was the training procedures described in [10] and the one employed to replicate the results in Table 1.

### 3.1. Dropout

In large neural networks it is difficult to cope with overfitting. An alternative to address this problem, without increasing the size of the database, is to apply the dropout scheme which temporarily deactivates units during training [21]. Fig. 4 shows a graphic description of this method.

### 3.2. Cross-validation

Overfitting may also be observed if the multilayer perceptron or DNN backpropagation-based training procedure is not stopped at the right time [22]. To avoid this problem, a cross validation method can be applied with a development set, i.e. after performing a given number of training epochs, the training is paused and the development set is propagated to compute the corresponding cost error. If the development set error cost increases or reaches a convergence criterion, the training is stopped and the neural network parameters of the previous epoch are retrieved. Otherwise, the training process is resumed. In this paper, the cross-validation scheme was performed in each training epoch.

In our SFM-DNN implementation, cross-validation was carried out with the CHiME-2 track 2 development set. The training was stopped if one of the two conditions was reached: the development error cost increases in 1% with respect the previous epoch; or, the development error cost decreases by less than 0.1% with respect the previous epoch.

### 3.3. Input-reference data normalization

As discussed in [23] [22], the neural network training procedure is sensitive to previous normalization (i.e. MVN or MN) performed to the input and reference features during the DNN training stage. Also, the results reported in [24] suggest that the utterance-by-utterance normalizations may give better results than, for instance, on a speaker-by-speaker basis. Consequently, we applied MVN to both the input and reference data on an utterance-by-utterance basis.

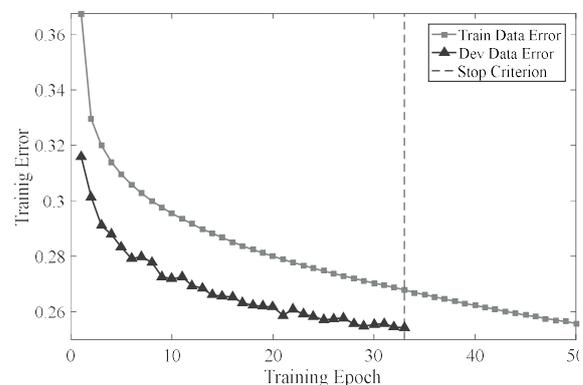

Figure 6: *SFM-DNN-E training error for both training and development databases, using cross-validation method. Vertical line indicates the stop criterion.*

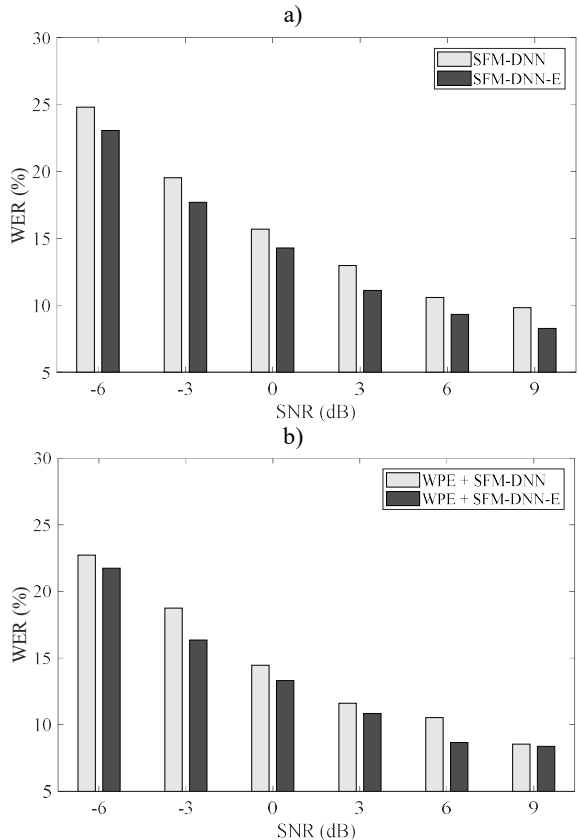

Figure 7: *WER (%) vs. SNR: a) SFM-DNN and SFM-DNN-E; and, b) WPE+SFM-DNN and WPE+SFM-DNN-E.*

### 3.4. Enhanced SFM-DNN

We incorporated dropout, cross-validation and utterance-by-utterance MVN normalization to the SFM-DNN training procedure. The WERs provided by the resulting system, SFM-DNN-E, are shown in Fig. 5. Figure 6 shows the evolution of the error cost with both the training and development data. The early stop of the training procedure should avoid the overfitting effect.

## 4. Discussion

According to Table 1, we were able to achieve similar results to those reported in [10] with SFM-DNN. Accordingly, SFM-DNN increased the WER in 0.8% when compared to the baseline system, i.e. noisy training, in conditions where there is less presence of additive noise and more presence of reverberation, i.e. 6 and 9 dB. As mentioned above, these results were our motivation to include a SE method, i.e. WPE, to reduce the effect of reverberation. As can be seen in Fig. 2, WPE presents a complementary behavior when compared to SFM-DNN with respect the improvement in recognition accuracy vs. SNR. As a result, the WPE+SFM-DNN system reduced the WER in 10.8% when compared to the baseline. However, WPE led to a higher WER reduction (11.4%) than WPE+SFM-DNN.

SFM-DNN was improved by incorporating modifications to the training procedure. As can be seen in Fig. 7, the resulting SFM-DNN-E system provided a WER that is 10.4% lower than the original SFM-DNN. When compared to WPE alone and the baseline system, WPE+SFM-DNN-E led to reductions in WER equal to 7.9% and 18.3%, respectively. The final average WER was 13.2% and compares very favorably with the one published elsewhere [12] [10] [13] [14] [15] with the same database, i.e. CHiME-2.

## 5. Conclusion

In this paper the effectiveness of a spectral feature mapping DNN is improved to remove the effects of reverberation and additive noise. Experiments with the CHiME-2 database show that this DNN is more effective in removing additive noise than reverberation. To counteract this limitation the DNN was combined with WPE that shows a complementary behavior with respect to the improvement in recognition accuracy vs. SNR. Also, modifications to the DNN training process were successfully applied. As a consequence, a final average WER equal to 13.2% and a relative reduction in WER as high as 18.3%, when compared to the baseline system with noisy training, were achieved. These results are very competitive with those published elsewhere using CHiME-2 database.

## 6. Acknowledgements

The research reported here was funded by Grants Conicyt-Fondecyt 1151306 and ONRG N°62909-17-1-2002. José Novoa was supported by Grant CONICYT-PCHA/Doctorado Nacional/2014-21140711.